# NASA-DARES: Request for Information NNH25ZDA002L

Topic 8: Prepare for the Discovery of Life Beyond Earth and Subsequent Post-Discovery Activities

## Collaborating with Artists in the Search for Life

Jack Madden - Blue Marble Space Institute of Science - jack.madden@bmsis.org
Cybele Collins - Department of Biological Sciences - University of Rhode Island
Mia Rollins - Department of Physics and Engineering - Community College of Rhode Island
Ashika Capirala - Department of Earth, Atmospheric, and Planetary Sciences - Purdue University

## Introduction

The draft report from the Communicating Discoveries in the Search for Life in the Universe (CDSLU) workshop outlines several ways collaboration between artists and scientists can unlock unique support for NASA's astrobiology press communication and public outreach.[1] The report's perspective on art focuses on how artists can make science more accessible to the public by harmonizing scientific accuracy, inspiration, and design. The report encourages broader participation and support for artist and scientist collaborations to meet the challenging communication goals of the field. While this approach is necessary to achieve the best possible outcomes in outreach and communication, there are critical means by which art can contribute to astrobiology beyond communication. We would like to argue for the further expansion of artist collaboration in astrobiology to include direct support of scientists and their goals through research-creation practices.

"Research-creation" is a method of using a conjunction of artistic and scientific processes to support the advancement of knowledge.[2] Viewing art merely as a form of advertisement for science through journal covers, educational programming, and creative storytelling limits the advantages science can draw from artists trained in creative problem solving, contextualization, and design-led research. Conceptual artists and designers trained in the rigorous development of ideas into form offer unique skills to address many of the problems scientists face. Research-creation may first appear as visualization of data or organizing scientists' stories, but goes beyond the confines of scientific tooling to help scientists grapple with the existential, moral, and social questions that arise from their work. The language within this NASA-DARES RFI solicitation admits that science research is inseparable from its societal impact and that scientists cannot operate as if in a romantic bubble of objective scientific solitude. Research-creation provides a parallel process of experimentation and feedback that can enhance our connection to and understanding of science through exploring these questions. To quote the solicitation, we must "consider and respond to the profound implications of life’s existence elsewhere, ensuring the field evolves to address new scientific, ethical, and societal changes": this is precisely the role of art.

The new approach and outcomes of research-creation are easiest to understand through examples. We will cover three examples, generalize their common traits, and suggest how support for such projects can be worked into the 2025 NASA-DARES.

### **Research-creation example 1:** Symposium on Habitability

In 1970, NASA commissioned the First National Symposium on Habitability to understand how long humans could survive in space. The event was organized by Dr. Ed Wortz, who worked on the physiological demands of astronauts, brought in psychologists, engineers, physicians, and social scientists to present their papers and find an answer. Dr. Wortz also invited Robert Irwin to present, who suggested that he host the symposium in his art studio instead. Irwin was a conceptual artist and knew his best contribution would be in his own medium of light and space, not words. Dr. Wortz agreed, and Irwin gathered other artists to help execute his plan to teach these experts a lesson in habitability.

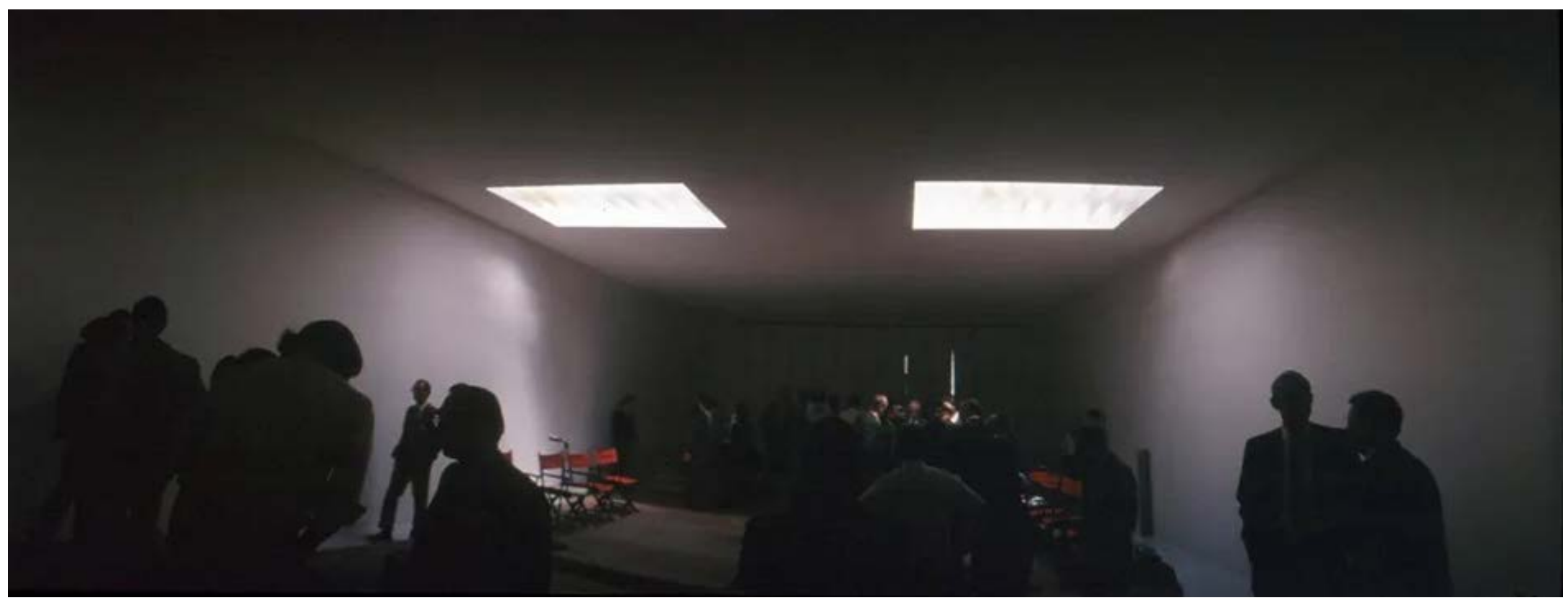

Figure 1. The first day of the NASA's habitability symposium. Credit: Larry Bell, 1970

On the first day, instead of the typical conference center, the throng of scholars wandered down an alleyway in Venice Beach, California, to find a literal hole in the wall that led them to their "habitat" for discussing long-term space travel (Figure 1). The room was dim and had just enough furniture to hold a basic conference. No outside light or sound could enter; they were in a capsule. Though uncomfortable, everyone was as determined as the astronauts they studied and carried on presenting about optimized sleep schedules and nutrition allotments.

The next day, the participants found a proper front door and the hall flooded with natural light from the newly uncovered windows. Everyone was a bit more relaxed and casual, and suit jackets started coming off as people opened up about their work.

On the third day, the entire front wall of the studio was gone. People on the street passed by and listened, fresh air came in, and the sound of waves set the tone (Figure 2). Serenity had taken over the symposium, and conversations became more honest, generous, and practical.

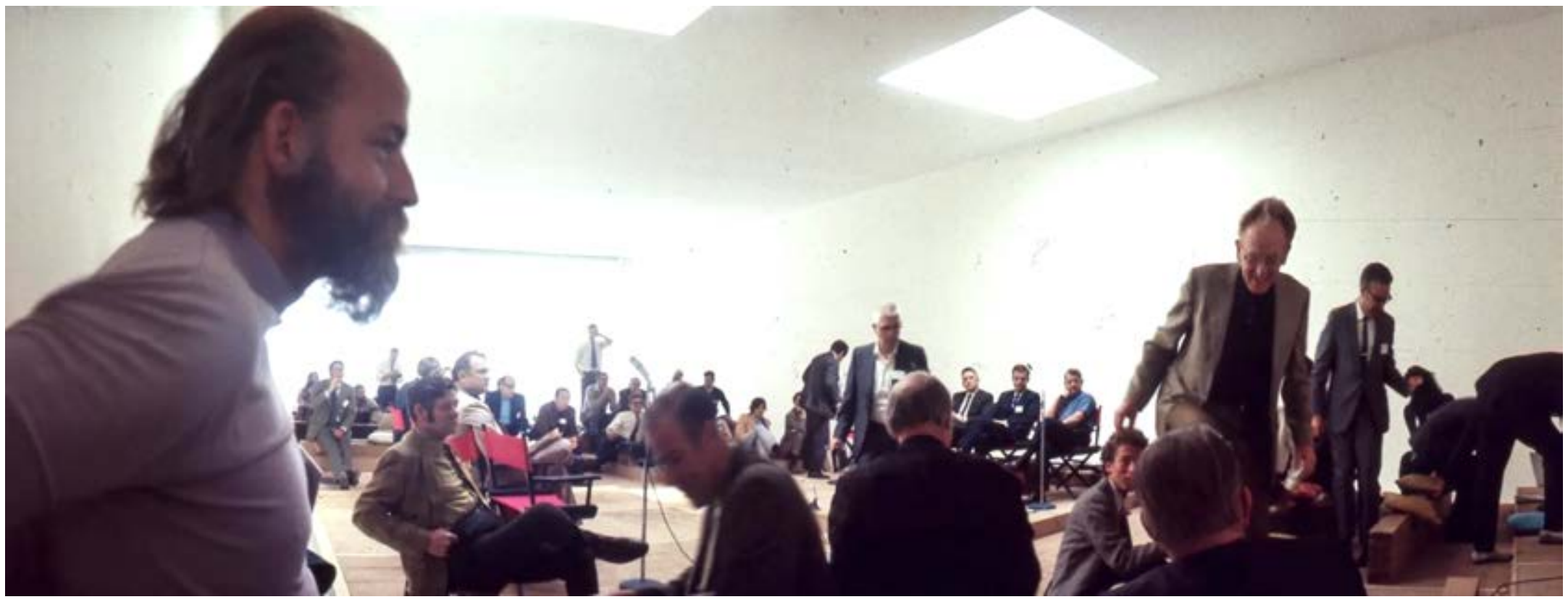

Figure 2. The third day of NASA's habitability symposium. Credit: Larry Bell, 1970

The attendees felt but did not consciously notice the changes being made. By the fourth day, there was enough buzz that they needed to discuss what had happened. The attendants soon realized that their environment significantly and unconsciously affected their mood and behavior. These experts began candidly accepting the shortcomings in their work and how habitability wasn't a technical hurdle an equation could solve. Windows for example, were considered too costly for a spaceship, now seemed essential for astronaut health. A new perspective on their work had materialized in 3 days without a single word from Robert Irwin, but they heard him loud and clear.

Afterward, Irwin said, "They tolerated this thing in the beginning—and then became aware that, essentially, this was an exercise in the problem they'd come to solve."[3]

At this point, the scientists and engineers in attendance had been designing Skylab, NASA's first space station, for more than three years. Thus, it was quite surprising to the NASA program director to receive a report, 10 days later, detailing 15 major design changes that would make the station more suitable for long-term missions.

Reflecting on the involvement of artists in science, Wortz, said. "There are things that need to be stated, things that need not so much to be communicated in the usual sense but brought out of people, things that are really there but need to be released. Enter the artist."[4]

Irwin and his colleagues were not the typical artists involved in science today. Their work is more likely found in a contemporary art museum than on the cover of Nature. They were conceptual artists who conducted experiments on experience and interaction to explore how humans process information into ideas. Their art process was much the same as the process of science. They rigorously studied the theory and results in their field well enough to profoundly contribute to the scientific community.

## **Research-creation example 2:** The Sagan Approach

Decades before the term "research-creation" was formulated, Carl Sagan was an avid practitioner of its approach to science. His authority as a career academic, presence in and around space missions, and humanist perspective led to some of the most well-known artistic interventions into astrobiology and perhaps all of science. The Voyager Golden Record, the Pale Blue Dot image, and the Cosmos series are all driven by research-creation in which Sagan simultaneously took on the role of conceptual artist and scientist. Their inspirational effect on non-experts is undeniable, but they also significantly impacted many scientists. The Golden Record project sparked new lines of inquiry for the participating scientists Philip Morrison, Bernard Oliver, Leslie Orgel, A. G. W. Cameron, Frank Drake, and many others.[5] Sagan's Pale Blue Dot image worked in harmony with his work on Galileo to rally scientists working in planetary science, environmental studies, climate change, and exoplanet research to think more holistically about the importance of their work (Figure 3).[6,7] Sagan's Cosmos series and work as a creative writer gave emerging and seasoned scientists alike a philosophical touchstone from which to contextualize and construct their ethic of exploration.

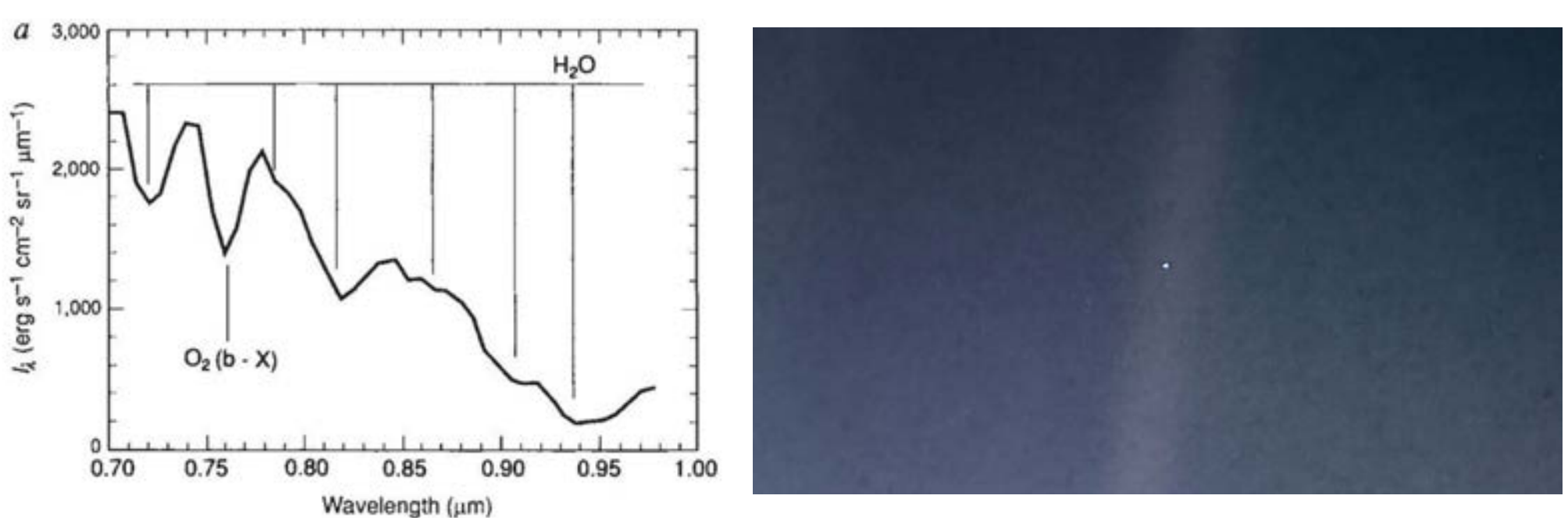

Figure 3. The Pale Blue Dot image and Earth's reflectance as seen from Galileo. These two complementary views made by Sagan show his scientific and artistic contributions to the complete picture of Earth's isolation as a habitable planet.

In astrobiology, contextualization significantly affects the questions scientists pursue and the future treatment of discovery. The search for life and its origins has garnered enormous scientific interest precisely because of the subjective relationships scientists form with the existential nature of the subjects. Research-creation interventions, like Sagan's, give scientists the opportunity to reflect upon that relationship. There is no built-in avenue for this reflection with the current academic system, but Sagan was able to produce these mirrors as an artist because he was invited in as a scientist. His type of research-creation needed active participation from scientific institutions and the openness from ranking scientists to see the cultural significance beyond the science products.

## **Research-creation example 3:** Transition Design

A promising framework for solving complex systems of problems called "Transition Design" is currently being pioneered out of Carnegie Mellon University.[8] The approach calls on designers, not administrators or subject matter experts, to play the central role in organizing plans for tackling system-wide socio-technical issues and focuses on visioning, theories of change, mindset, and new design. The Transition Design framework gives designers the tools to work with all stakeholders to build a new system of narratives, pathways, and goals toward the most desirable outcome. This can be particularly helpful in aligning interests when scientists expose paradigm-shifting results that would lead to vast socio-political adjustments, such as with climate change. By applying the principles of Transition Design to scientific problems, it is clear that there should be room made within the scientific method to include steps for humility, reflection, cross-disciplinary communication, and respect for broader implications if we are to find more efficient and satisfying outcomes for our biggest open questions.

Principles from the Transition Design framework could help NASA Astrobiology guide community involvement in creating a Confidence of Life Detection (CoLD)-like scale that sees wide adoption[9], prepare post-detection protocols for data release and communicating between scientists, make sense of potential technosignatures, and seamlessly integrate ethical and societal considerations into mission planning. For example, many astronomers may be in a position to find life but have gone through no academic training to learn how to address the ethical implications of their work. A Transition Design approach could explore how this oversight would impact the future of the field and identify ways of carefully implementing ethical research standards into Astronomy to better align with fields like Biology.

## Summary

These examples highlight three of the many benefits of including a research-creation approach along with a scientific project.

**Creative problem solving:** In the habitability symposium, artists designed an intervention to break scientists outside their traditional work patterns and see problems from novel perspectives. Scientists are trained to see problems through theory or experimental lenses, while big breakthroughs may be hidden on a mindset or posture level, areas where conceptual artists are uniquely trained to offer creative solutions.

Other examples: Jon Lomberg's painting of the Milky Way, which led to new insight into galactic structure; Erika Blumenfeld's work cataloging samples in NASA's astromaterials lab; and Daniela de Paulis's creation of a mock SETI scenario.

Suggested implementation: Include language in grant awards that expressly sets funds aside for artist consultation in research projects with guidelines on productive research-creation practices. Clarify how research funds within existing grants and fellowships can be used to consult artists.

**Contextualization:** Carl Sagan generated contextual fuel that helps scientists align their values with their work. Moral reflection and meaning creation are essential for scientists to function as valuable members of the society their discoveries impact. There is no greater need than in the case of life detection for scientists to proceed ethically and with authentic awareness of the profound impact their work may release. Having measures in place to allow for Sagan-like artistic interventions is a key part of bringing scientists face-to-face with the context of their work while simultaneously helping reconnect scientists with non-experts.

Other examples include director Webb's efforts at NASA in the 1960s to provide artists with access to the Apollo program, photographer Richard Underwood teaching astronauts to shoot the Blue Marble and Earthrise photos, Noll and Bond's Hubble Heritage Project, and the SETI Institute's artist-in-residence program.

Suggested implementation: Reignite the NASA art program with artist-in-residence fellowships and provide "in-the-room" access for artists and scholars to study missions and astrobiology initiatives. Grants could be made for artists to document astrobiology research and collaborate with researchers, similar to what was done for the Apollo program.

**Design-led research:** New approaches to planning science, such as Transition Design, can bring scientific endeavors closer to their intended outcome by injecting moments of reflection and spawning productive collaboration between connected disciplines. Incorporating designers in planning scientific experiments and field-level strategies can make reaching goals more efficient and reveal unseen connections between fields.

Other examples include Edward Tufte's contributions to improving scientific visualization design, the team of designers at The Studio within JPL, and the integration of design into science research at the MIT Media Lab.

Suggested implementation: Plan life discovery experiments, CDSLU workshops, and CoLD scale implementation using transition design frameworks to account for all stakeholder input. Create a version of JPL's Studio dedicated to creative problem solving for the astrobiology community. This studio could coordinate artist grant programs and provide central support for the field's research-creation initiatives.

As the first caretakers of new knowledge, scientists have a fundamental responsibility over how we understand our world. We hope there is enough support for the scientists within the astrobiology community who are willing to embrace research-creation techniques and, in so doing, better equip all of us to make profound discoveries. The future NASA should create for astrobiology could include Sagan-like interventions within the Habitable Worlds Observatory mission plan, astronomers working with art historians to recognize art forms among technosignatures, or biologists and architects discussing missing links in building life. To understand the merit of these futures, the NASA DARES task force needs a nuanced approach to "art" that distinguishes between "communication-based" and "research-creation." The search for life and its origins will take us down paths stranger than scientists can predict; we must invite those trained in turning imagination into reality; we must invite artists.